\documentclass[twocolumn]{aastex62}
\usepackage[para]{threeparttable}
\usepackage{xcolor}
\usepackage[caption=false]{subfig}

\received{January 3, 2019}
\revised{}
\accepted{July 15,2019}
\submitjournal{AJ}

\shorttitle{sodium in exoplanetary atmospheres}
\shortauthors{Zak, Kabath, Boffin et al.}

\begin{document}

\title{High-resolution transmission spectroscopy of four hot inflated gas giant exoplanets}

\correspondingauthor{J. \v{Z}\'{a}k}
\email{437396@mail.muni.cz}

\author[0000-0001-9416-9007]{Ji\v{r}\'{i} \v{Z}\'{a}k}
\affiliation{Astronomical Institute, Czech Academy of Sciences, 
              Fri\v{c}ova 298, 25165, Ond\v{r}ejov, Czech Republic}
\affiliation{Department of Theoretical Physics and Astrophysics, Masaryk University, Kotl\'{a}\v{r}sk\'{a} 2, 61137 Brno, Czech Republic
}

\author[0000-0002-1623-5352]{Petr Kabath}
\affil{Astronomical Institute, Czech Academy of Sciences, 
              Fri\v{c}ova 298, 25165, Ond\v{r}ejov, Czech Republic}


\author[0000-0002-9486-4840]{Henri M.J. Boffin}
\affiliation{ESO, Karl-Schwarzschild-str. 2, 85748 Garching, Germany}

\author[0000-0002-5963-1283]{Valentin D. Ivanov}
\affiliation{ESO, Karl-Schwarzschild-str. 2, 85748 Garching, Germany}

\author[0000-0002-7602-0046]{Marek Skarka}
\affil{Astronomical Institute, Czech Academy of Sciences, 
              Fri\v{c}ova 298, 25165, Ond\v{r}ejov, Czech Republic}
\affiliation{Department of Theoretical Physics and Astrophysics, Masaryk University, Kotl\'{a}\v{r}sk\'{a} 2, 61137 Brno, Czech Republic
}



\begin{abstract}
The technique of transmission spectroscopy allows us to constrain the chemical composition of the atmospheres of transiting exoplanets. It relies on very high signal-to-noise spectroscopic (or spectrophotometric) observations and is thus most suited for bright exoplanet host stars. In the era of TESS, NGST and PLATO, more and more suitable targets, even for mid-sized telescopes, are discovered. Furthermore, a wealth of archival data is available that could become a basis for long-term monitoring of exo-atmospheres.
We analyzed archival HARPS spectroscopic time series of four host stars to transiting bloated gas exoplanets, namely WASP-76b, WASP-127b, WASP-166b and KELT-11b, searching for traces of sodium (sodium doublet), hydrogen (H$\alpha$,\ H$\beta$), and lithium (670.8 nm).
The archival data sets include spectroscopic time series taken during transits. Comparing in- and out-of-transit spectra we can filter out the stellar lines and investigate the absorption from the planet. Simultaneously, the stellar activity is monitored using the Mg I and Ca I lines.
We independentely detect sodium in the atmosphere of WASP-76b at a 7--9\,$\sigma$ level. Furthermore, we report also at 4--8\,$\sigma$ level of significance the detection of sodium in the atmosphere of WASP-127b, confirming earlier result based on low-resolution spectroscopy. The data show no sodium nor any other atom at high confidence levels for WASP-166b nor KELT-11b, hinting at the presence of thick high clouds. 
\end{abstract}



\section{Introduction}

Transmission spectroscopy has the power to probe the physical conditions in
the atmospheres of exoplanets, as it allows detecting the various  absorption features that are imprinted on the
stellar light that passes through their atmospheres. Not surprisingly, the
first detections of features in exo-atmospheres were made from space
\citep{charb03,vidal2003}, while ground-based transmission spectroscopy delivered its
first secure detection of sodium in extrasolar atmospheres only about a decade ago
\citep{Redfield2008,Snellen2008}.  If from space, only low-resolution spectroscopy can be used, from the ground, both low-resolution spectrophotometry and high-resolution spectroscopy can and have been used to detect atoms and molecules in the atmospheres of several exoplanets. In particular, in most recent years, high-resolution spectroscopy has been more and more successful. Moreover, \cite{wytt15} used HARPS at the ESO 3.6-m telescope to be able to detect
sodium in HD\,189733b, demonstrating that spectrographs attached to moderate-size telescopes
can also be used for this purpose. In the past only a handful of suitable, bright enough targets were
available for such kind of study, but this is now changing rapidly with the discoveries from
the {\it TESS} mission \citep{TESS,kabath}.
The high resolving power, i.e., 
R$\geq$40\,000--60\,000, removes the need for a reference star that is at the same
time close on sky to the target, and of similar brightness -- two rarely met
conditions that are limiting the low-resolution spectrophotometry studies. In
contrast, the high resolving power solely requires accurate telluric correction and
a careful analysis of the radial velocities of the star-planet system \citep[e.g.,][]{rojo,wytt17}.

 Another advantage of high-resolution spectroscopy is its ability to resolve and characterize individual planetary absorption lines. In particular, the 
Na D lines are well suited for transmission spectroscopy due to the strong expected signal and the fact that they are very narrow features, in comparison to the broad molecular absorption bands of, e.g., $\rm{CH_4,\ H_2 O,\ TiO\ and\ VO.}$

If sodium, and in some cases, potassium measurements
based on low-resolution spectra were reported
\citep[e.g., for some of the most recent ones,][]{2016A&A...596A..47S,Chen2017,palle,Murgas2017,Lendl2017,palle2,pearson}, only a few studies detected sodium using high-resolution transmission spectroscopy:
\cite{wytt17} registered strong absorption signal in WASP-49b.
Using HARPS-N mounted on the TNG, \cite{casa} and \cite{casa18}  concluded to the presence of sodium in the atmospheres of WASP-69b and MASCARA-2b. \cite{khala} tentatively detected sodium in WASP-17b using the MIKE instrument on the 6.5m Magellan Telescope. Finally, \cite{jensen} detected  sodium and hydrogen in WASP-12b using the Hobby-Eberly Telescope and \cite{deibert} found sodium in HAT-P-12b using the Subaru Telescope.

These are not the only elements to have been discovered in the atmospheres of exoplanets, as there are published detection of hydrogen \citep{Ehrenreich2008,Lecavelier2010}, helium
\citep{Nortmann2018,Spake2018}, calcium, scandium \citep{rojo}, TiO and VO
\citep{Deseret2008,seda}, CO \citep{Snellen2010}, H$_2$O and CH$_3$
\citep{Tinetti2007,Swain2008,Birkby2013}. Some of these detections were disputed by
\citet{Gibson2011} who discusses in great detail the complex systematics that affect
the results, even those obtained from space. This issue was addressed in recent
years by paying closer attention to the systematic effects \citep[e.g.,][]{boffin} and by
introducing new stable instruments. 

It is also important to note that there is a tremendous amount of archival spectra -- obtained during transits for other purposes, such as measuring the spin-orbit alignment measurements via the Rossiter-McLaughlin effect \citep{Queloz2000} -- that are available for further analysis. They can set the foundation of long-term exoclimate studies. Therefore, the method of high-resolution transmission spectroscopy from the ground is likely to become even more widely used than today, and to gain further importance for the characterization of exoatmospheres.

  \begin{figure*}{}
    \resizebox{\hsize}{!}
             { \includegraphics[bb=5 5 800 500,clip]{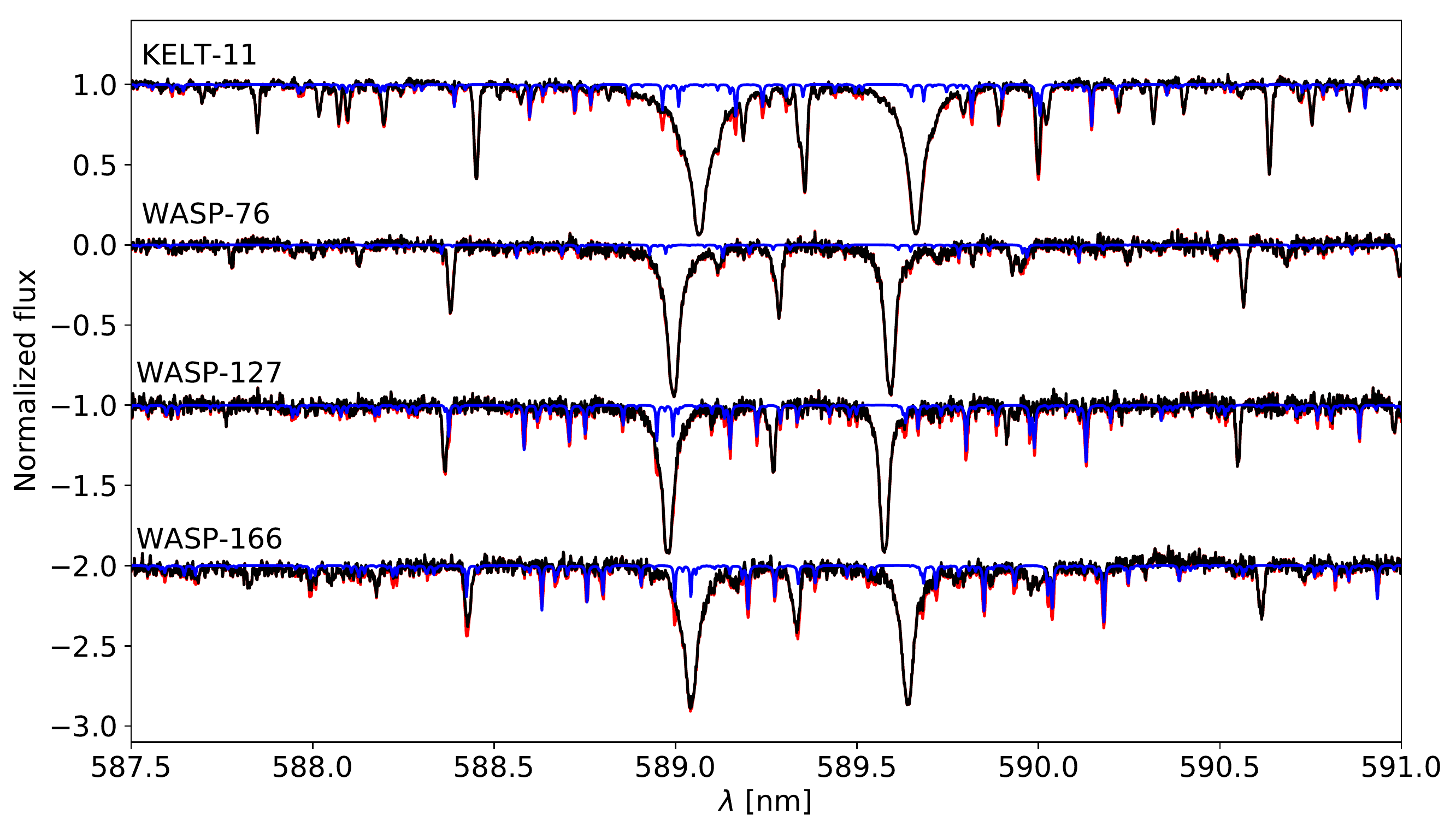}}
       \caption{Illustration of the spectral region around the Na D lines for one selected example spectrum for each of our targets. The observed spectra in red are presented in the stellar reference frame during mid-transit. Telluric spectra are shown in blue. Observed spectra after telluric correction are presented in black.
                }
          \label{f1}
    \end{figure*}

Here we investigate the atmospheric chemistry of two inflated super-Neptunes and two inflated hot-Jupiters with archival HARPS data\footnote{Data were obtained under Programme IDs: 090.C-0540(F), 098.C-0304(A) and 0100.C-0750(A)}. We study the instrument systematics that can influence the results and the effects of the stellar activity \citep[see][for further discussion]{klocova} using the calcium lines as activity indicators \citep{kh17}. This work is part of a larger effort to characterize systematically and uniformly a large number of exoplanetary atmospheres, and to shed more light onto their chemical composition. We also aim to identify interesting targets for CHEOPS and JWST follow-up.

\section{Data sets and their analysis}

\subsection{Data summary}
Table~\ref{obs_logs} shows the various datasets we used, while Table~\ref{tab:targets} presents the main properties of our targets. For illustration purpose we show part of the spectrum around the sodium Na D region of each of our targets, along with the telluric spectrum in Fig. \ref{f1}.\\

\begin{table*}[htbp]
\begin{center}
\caption{Properties of the four targets (star and planet) considered in this paper.}
\begin{tabular}{@{ }l@{ }l@{ }c@{ }c@{ }c@{ }c@{ }} 
\hline
& Parameters & WASP-76  & WASP-127  & WASP-166 & KELT-11 \\ 
\hline
Star & V mag & 9.5 & 10.5 & 9 & 8 \\
& Sp. Type & F7 & G5 & F9 & G8/K0 \\
& M$_s$ (M$_\odot$) & 1.46 $\pm$ 0.07 & 1.31 $\pm$ 0.05 & 1.19 $\pm$ 0.06 & $1.438^{+0.061}_{-0.052}$ \\
& R$_s$ (R$_\odot$) & 1.73 $\pm$ 0.04 & 1.33 $\pm$ 0.03 & 1.22 $\pm$ 0.06 & $2.72^{+0.21}_{-0.17}$ \\
& T$_{eff}$ (K) & 6250 $\pm$ 100  & 5750 $\pm$ 100 & 6050 $\pm$ 50  & $5370^{+51}_{-50}$ \\
& log g & 4.4 $\pm$ 0.1 & 3.9 $\pm$ 0.1 & 4.5 $\pm$ 0.1 & $3.727^{+0.04}_{-0.046}$ \\

Planet & M$_p$ (M$_{\rm Jup}$) & 0.92 $\pm$ 0.03 & 0.18 $\pm$ 0.02 & 0.102 $\pm$ 0.004 & $0.195^{+0.019}_{-0.018}$ \\
& R$_p$ (M$_{\rm Jup}$) & $1.83^{+0.06}_{-0.04}$  & 1.37 $\pm$ 0.04 & 0.63 $\pm$ 0.03 & $1.37^{+0.15}_{-0.12}$ \\
& Period (d) & 1.809886 $\pm$ 
0.000001 & 4.178062 $\pm$ 0.000002 & 5.443526 $\pm$ 0.00001  & $4.736529^{+0.000068}_{-0.000059}$ \\
&Transit duration (days) &0.1539 $\pm$ 0.0008 &0.1795 $\pm$ 0.0007&0.148 $\pm$ 0.008 &$0.3051^{+0.0053}_{-0.0051}$\\
&Orbital Semi-major axis (AU)&0.33 $\pm$ 0.0005 &0.052 $\pm$ 0.0005& 0.0642 $\pm$ 0.0001&$0.06229^{+0.00088}_{-0.00076}$\\
&Orbital inclination (degrees)&$88.0^{+1.3}_{-1.6}$  &$88.7^{+0.8}_{-0.6}$&87.8 $\pm$ 0.6& 	$85.8^{+2.4}_{-1.8}$ \\
& Reference & West et al. (2016) & Lam et al. (2017) & Hellier et al. (2018) & Pepper et al. (2017)\\
\hline 
\end{tabular}
\label{tab:targets}
\end{center}
\end{table*}

\begin{table*}[htbp]
\begin{center}
\caption{Observing logs for all four data sets. Number in parenthesis represents number of frames taken in-transit.$^1$ The signal to noise rate, S/N, is derived from the 16th order (585.03-591.62~nm) of the echelle spectrum, corresponding to the position of the sodium doublet line.}
\begin{tabular}{cccccc}
\hline
Date & No.     & Exp.     & Airmass & Median & S/N$^1$ \\ 
     &~Spectra~&~Time [s]~& range   &  S/N   & \\ 
\hline
\multicolumn{5}{c}{} \\
\multicolumn{5}{c}{WASP-76} \\
2012-11-12~&~64 (40)~& 300     & 1.79-1.18-1.48~&~21.1-31.6&13.3-20.6 \\  
2017-10-25~&~49 (27)~& 400-600 & 1.75-1.18-2.10~&~32.1-55.8 &18.7-34.2\\
2017-11-23~&~66 (40)~& 300-400 & 1.42-1.18-2.45~&~30.7-53.5& 15.5-31.8 \\ 
\multicolumn{5}{c}{} \\
\multicolumn{5}{c}{WASP-127} \\
2017-02-28~&~37 (22)~& 500     & 1.40-1.11-1.47~&~29.9-40.0&14.7-20.1 \\  
2017-03-20~&~45 (30)~& 500-600 & 1.98-1.11-1.55~&~26.4-38.2 & 14.0-19.5\\
\multicolumn{5}{c}{} \\
\multicolumn{5}{c}{WASP-166} \\
2017-01-14~&~75 (39)~& 300-350 & 2.37-1.01-1.17~&~20.0-52.8&11.0-33.0 \\  
2017-03-04~&~52 (39)~& 300-400 & 1.01-1.21-2.55~&~26.4-65.0&13.5-37.3 \\
2017-03-15~&~66 (34)~& 350     & 1.17-1.01-2.35~&~18.8-55.3& 8.6-29.9 \\ 
\multicolumn{5}{c}{} \\
\multicolumn{5}{c}{KELT-11} \\
2017-02-02~&37 (24)~& 300     & 1.42-1.11-1.08~&~81.2-92.8 &45.6-53.4 \\  
2017-02-15~&28 (0) ~& 400     & 1.11-1.08-1.25~&~85.4-106.8&45.4-59.2 \\
2017-02-16~&69 (49)~& 400     & 2.29-1.06-1.93~&~46.1-115.3&24.9-64.8 \\ 
2017-02-17~&27 (0) ~& 400     & 1.81-1.23-1.07~&~67.5-95.4&39.0-55.5  \\  
2017-03-06~&42 (0) ~& 300     & 1.06-1.21-2.04~&~26.3-89.6&11.1-43.4  \\
2017-03-07~&93 (71)~& 300     & 2.23-1.06-2.41~&~48.6-83.1&26.1-45.5  \\ 
2017-03-08~&46 (0) ~& 300     & 1.60-1.12-1.09~&~70.1-91.3&38.4-50.2  \\
\hline 
\end{tabular}

\label{obs_logs}
\end{center}
\end{table*}

\noindent {\bf WASP-76b}:
The HARPS data are covering 3 transit events, on 11/12 November 2012, 24/25 October 2017 and 22/23 November 2017. The total number of frames was $176$ with $105$ in-transit, and the exposure times were  between 300-600 seconds. 
\\


\noindent {\bf WASP-127b}:
The HARPS data are covering 2 transit events during the nights of 27/28 February and 19/20 March 2017. The total number of frames was $82$, of which 52 were in-transit. 
\\


\noindent {\bf WASP-166b}:
The HARPS data are covering three transits during the nights of 13/14 January, 03/04 March and 14/15 March 2017. The total number of frames was $188$, of which 112 were in-transit. 
\\


\noindent {\bf KELT-11b}:
The HARPS data are covering 3 transit events and 4 nights of out-of-transit on 01/02 February (transit), 14/15 Februray, 15/16 February (transit), 16/17 February, 05/06 March, 06/07 March (transit) and 07/08 March 2017. The total number of frames was 342, of which 144 were in-transit. 
 \\

\subsection{Data Reduction}

We retrieved the publicly available HARPS \citep{harp} reduced spectroscopic data sets from the ESO Science data archive for the four planets WASP-76b, WASP-127b, WASP-166b and KELT-11b (see Table \ref{tab:targets}). The retrieved data are fully reduced products obtained with the HARPS Data Reduction Software (DRS version 3.5). 
In more detail, each spectrum is provided as a merged 1D spectrum with 0.01 $\AA$ wavelength step. The reduced spectrum covers the wavelength range between 380 nm and 690 nm, and has a spectral resolution R~$\approx$~115\,000 corresponding to 2.7~km\,s$^{-1}$ per resolving element. Spectra are corrected to the barycentric frame of reference. The sodium doublet is well-centered on the 16th order of HARPS.

\subsection{Normalization}
{We obtained the radial velocities of the star (see below) from the FITS header of the spectra and subsequently shifted both during transit $f(\lambda,t_{\rm{in}})$ and out of transit $f(\lambda,t_{\rm{out}})$ frames to the stellar rest frame. We have made a cut of the region of interest and performed a normalization for the obtained spectra $f(\lambda,t_{\rm{in}}) \rightarrow \tilde{f}(\lambda,t_{\rm{in}})$ and $f(\lambda,t_{\rm{out}}) \rightarrow \tilde{f}(\lambda,t_{\rm{out}})$, where the tilde represents normalization. Normalization has been carried out using the standard IRAF routine \textit{continuum}. More precisely, a Legendre polynomial has been used to perform the normalization of continuum regions without strong lines within the regions centred around the Na D lines (583 to 596 nm), H$\alpha$\  (645 to 670 nm), H$\beta$\  (475 to 495 nm) and lithium (666 to 674 nm).

\subsection{Telluric correction}
As the next step, the Earth's atmospheric signature needs to be accounted for. There are several ways to remove telluric lines from the spectra.  One of the methods is to simultaneously observe an early type rapidly rotating star to obtain the telluric spectrum. However, this method has a disadvantage of reducing the observing time spent on the primary target. Another method is described in \cite{rojo} and \cite{wytt15} where they use the fact that the variation of the telluric lines depth follows the airmass variation linearly. We decided to follow the method used in \cite{casa}. We took the one-dimensional telluric spectrum constructed from the line list of HITRAN \citep{rot} and matched its resolution to the resolution of the data. Then we corrected the reference telluric spectrum for the radial velocity to the same frame as the observed spectrum. The telluric lines absorption depth is changing with the airmass. Thus we used a Least Square Method on the unblended telluric lines to scale the telluric model to the same airmass (depth) as the observed spectrum. Afterwards, we divided each observed spectrum by the telluric spectrum scaled for the given airmass using standard IRAF commands resulting in the removal of the telluric lines. As discussed in \cite{casa}, the scaling of the telluric spectra is not perfect as line depth variation is not equal for all the telluric lines, hence possibly introducing small residuals into the final transmission spectrum. As an illustration,
the RMS of a single-night transmission spectrum in the telluric region (589.9 to 590.4 nm) before applying the telluric correction is 0.00339. The RMS of the same region was decreased to 0.00196 after applying the correction. For comparison, the RMS of the spectral region without telluric features (587 to 587.5 nm) is 0.00193.
\subsection{Transmission spectrum}\label{sec:Transmission_spectrum}

After the telluric correction, we averaged all out of transit spectra to create a ``master out'' frame $\tilde{F}_{\rm out}(\lambda)=\sum_{\rm{out}} \tilde{f}_{\rm{out}}(\lambda,t_{\rm{out}})$. Next, we divided each in-transit frame $\tilde{f}_{\rm in}(\lambda,t_{\rm{in}})$ by $\tilde{F}_{\rm out}(\lambda)$ to create the individual transmission spectra. Each of these frames was subsequently corrected for the planetary motion as the planetary signal is shifted from the stellar reference frame. We created a model of radial velocity taking only out-of-transit data\footnote{As the planetary signal is shifted by as much as tens of kilometers from the rest frame, using the observed radial velocities would imply correcting for the possible Rossiter-McLaughlin effect and would produce systematic effects in the final transmission spectrum \cite{cegla, wytt17}.}. Finally, we summed up all the individual transmission spectra and normalized this sum to unity and then subtracted unity to retrieve the final transmission spectrum, $\tilde{\mathcal{R}}(\lambda)$:


\small
\begin{center}
\begin{equation} \label{trans}
\tilde{\mathcal{R}}(\lambda)=\sum_{\rm{in}} \left. \frac{\tilde{f}_{\rm{in}}(\lambda,t_{\rm{in}})}{\tilde{F}_{\rm{out}}(\lambda)}\right|_{\rm{Planet\ RV\ shift}} -1
\end{equation}
\end{center}
\normalsize

\subsection{Error estimates}
 
The uncertainties of the exoatmospheric Na D lines measurements
were estimated with a simple simulation from the data itself. This
approach has the advantage of preserving as much as possible the
systematic errors of the data.

The starting point of the simulation is our actual measurement --
the fit that reproduces the observed spectrum with a flat line and
two Gaussians (see Sec.\,\ref{sec:Transmission_spectrum}). In the
following we only consider a region of the spectrum containing 190
pixel elements, centred on the Na doublet. We converted the fit
within this region into a noiseless spectrum with the same
wavelength sampling as the original observational data.

The first step of our error analysis is to calculate the residuals
of the observational data with respect to the fit. Second, the
residuals were shifted by one pixel and then added to the best fit
model spectrum. Thus, the residual from the first pixel of the data
was added to the second pixel of the fit, and the residual from
the last pixel of the data was added to the first pixel of the best
fit. This created a new realization of artificial data that have
the same Na D signal and a different noise for each pixel. However,
the overall statistical properties of the noise are preserved: for
example if there is a systematic trend of increasing noise toward
the end of the observations, because of increasing airmass, it will
be preserved. The third step is to remeasure the Na absorptions in
the new artificial spectrum.

The last two steps were repeated 189 times, creating 189 unique new
realizations -- each time the shift of the residual pattern was
increased by one, so for the second realization the first deviation
was added to the third pixel and the last deviation was added to
the second pixel. For the third realization the first deviation was
added to the fourth pixel and the last deviation --~to the third
pixel. In general, to create the i$^{\rm th}$ realization we added
the first deviation to the (i+1)$^{\rm th}$ pixel of the noiseless
model spectrum and the last realization -- to the i$^{\rm th}$
pixel of the noiseless model spectrum. Summarizing, this exercise
is a simple shift of the residuals in a circular manner. It has
been used before, e.g. by \citet{caceres09,caceres11}.

The measurement errors were derived from the distributions (over
all realizations) of the fitting parameters determined for each
realization in the third step of the process described above.
These distributions are shown in Fig.\,\ref{boot}. We adopt as
final measurements and as their errors the means and the FWHMs
values of these distributions. 

\begin{figure}[htp]
\subfloat[WASP-76]{%
  \includegraphics[clip,width=\columnwidth]{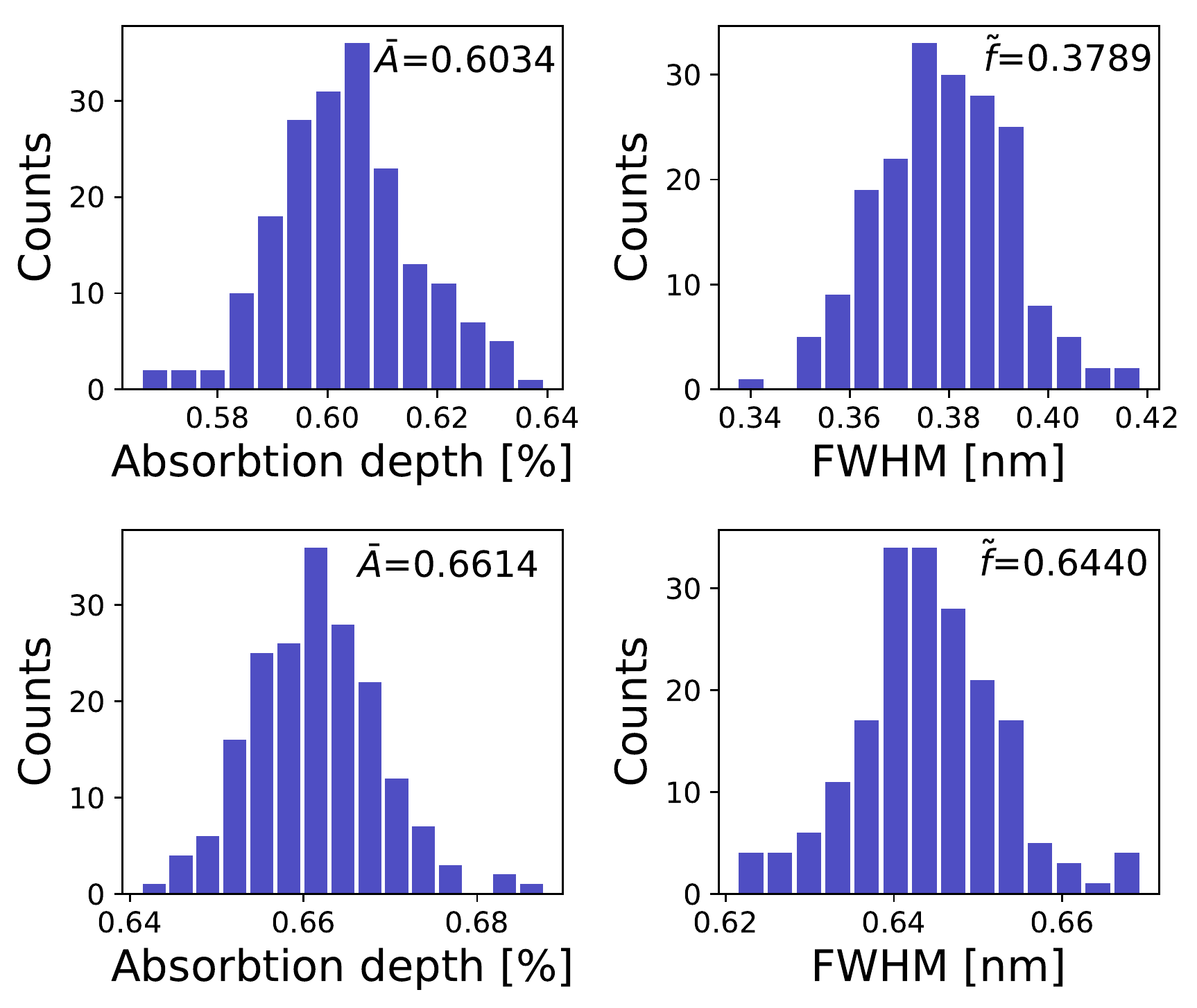}%
}

\subfloat[WASP-127]{%
  \includegraphics[clip,width=\columnwidth]{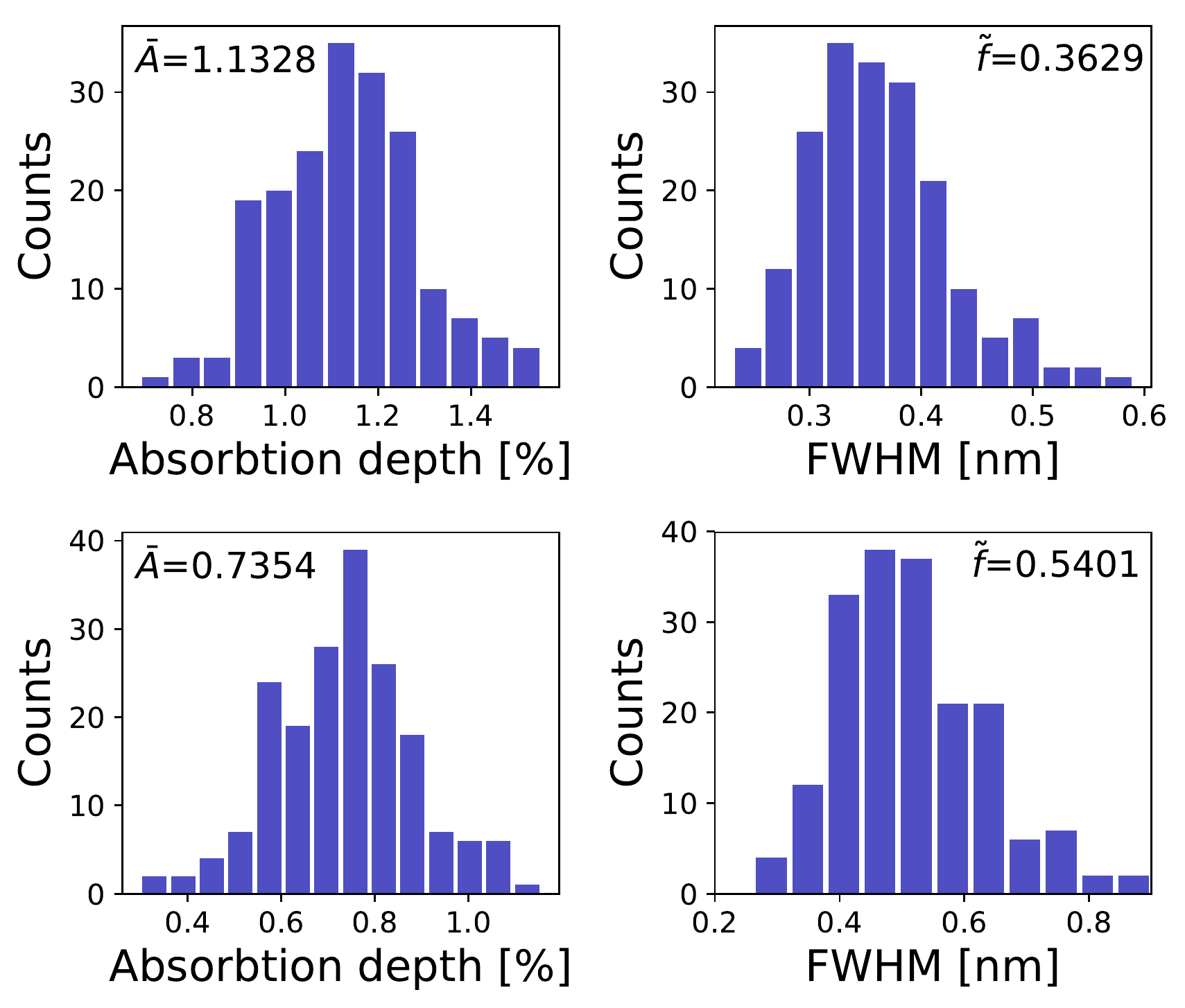}%
}

\caption{Distribution of the fitting parameters. The top rows of each subfigure show the parameter distribution for the D2 line while the bottom rows show the distribution for D1. Each plot displays the mean value. See text for a description of the method.}
\label{boot}
\end{figure}
 
%
%

\subsection{Stellar activity}

We also checked for possible false-positive detections. As suggested in \cite{barn}, stellar activity could produce signatures in lines originating mainly from the stellar chromosphere. To rule out any such fake signals we investigated the vicinity of Mg I (581.36\,nm) and Ca I (612.22 and 616.22 nm) lines. The lack of any features in these regions shows that the increase of absorption in the sodium Na D lines is not caused by the chromospheric activity or by systematic effects in our analysis.

 \begin{figure*}[htbp]
    \resizebox{\hsize}{!}
            { \includegraphics[width=\hsize]{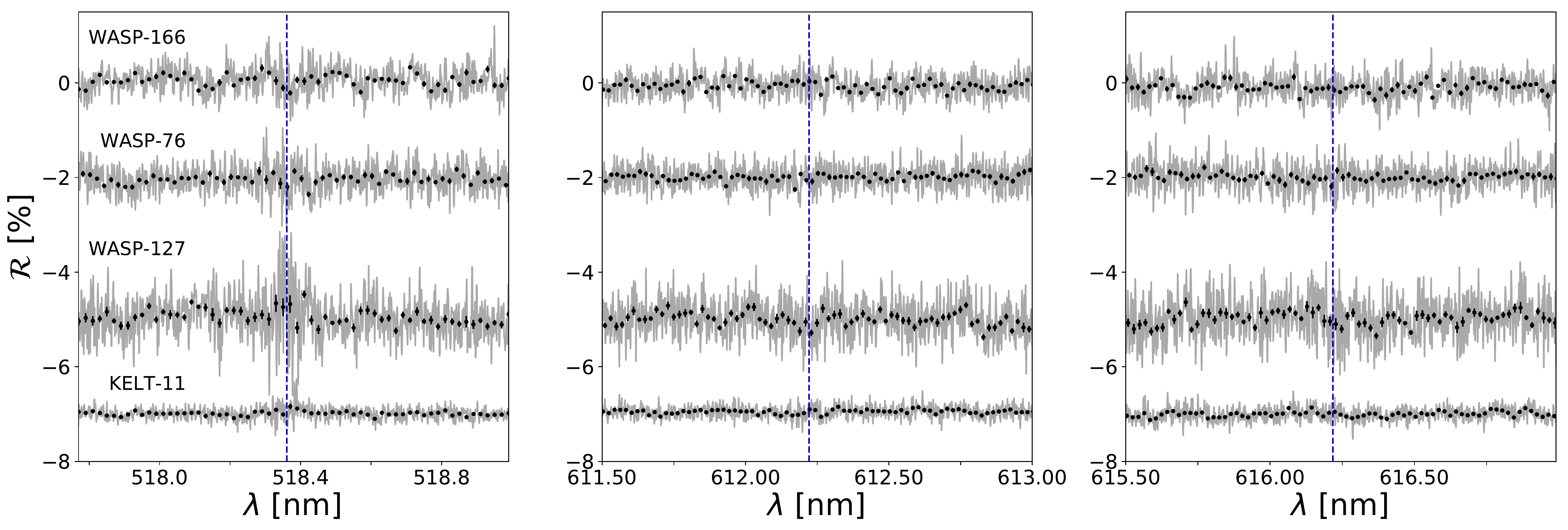}}
      \caption{Activity cross-check region of Ca I and Mg I lines for all our targets. From left to right, figure shows Mg I and both Ca I regions. Blue dashed lines indicate the exact position of Mg I (518.3604 nm) and Ca I (612.2217 nm a 616.2173 nm) lines. Vertical offset was made for better readability.}       
         \label{w76bact}
   \end{figure*}

\subsection{Rossiter--McLaughlin effect}\label{rmef}

When looking at the presence of any atomic or molecular species in the spectrum of a planet, it is important to take into account the Rossiter--McLaughlin effect \citep[R-M;][]{Queloz2000,Ohta2005}, as shown by \citet{wytt17}. In Fig.~\ref{fig:RM}, we show the measured radial velocities of our targets as obtained with HARPS as well as the expected orbital motion of the star due to the presence of the planet. Any deviation from this could be due to the R-M effect. This is indeed the case for WASP-166, where the R-M effect is clearly visible (as also shown by \cite{helier18}). 
For the three other targets, the R-M effect is not measurable.
It is possible to have an order of magnitude estimate of the R-M effect \citep{2018haex.bookE...2T}:

\begin{center}
\begin{equation} \label{rm}
\Delta v_r \simeq \frac{2}{3} \gamma^2 V \sin i \sqrt{1-b^2},
\end{equation}
\end{center}

where $V \sin i$ is the rotational velocity of the star, $b$ is the impact parameter and $\gamma = \frac{R_p}{R_s}$, with $R_p$ being the radius of the planet and $R_s$ that of the star.
Estimating these values for the 4 targets we get the following values for $\Delta v_r$: 2.3 m\,s$^{-1}$ (WASP-76),  2 m\,s$^{-1}$ (WASP-127), 8~m\,s$^{-1}$ (WASP-166), and 4 m\,s$^{-1}$ (KELT-11). These are in agreement with what we see in Fig.~\ref{fig:RM}.

Moreover, in all cases, the R-M effect is too small to have any real impact on the depth of the Na D lines \citep{Nortmann2018}. We thus conclude here that the R-M effect is not important in our analysis.

 \begin{figure*}
   \centering
   \includegraphics[width=0.45\hsize]{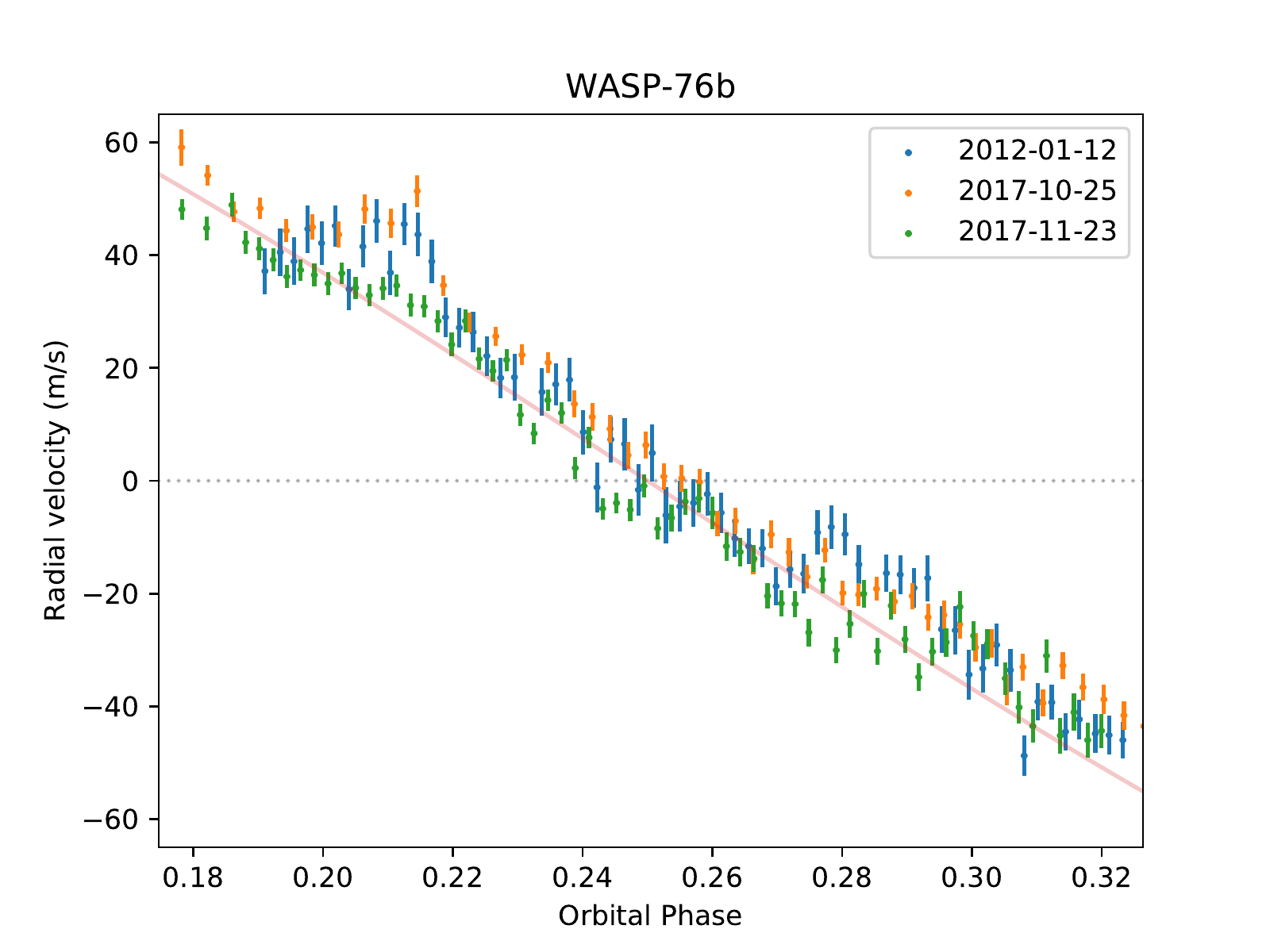}
    \includegraphics[width=0.45\hsize]{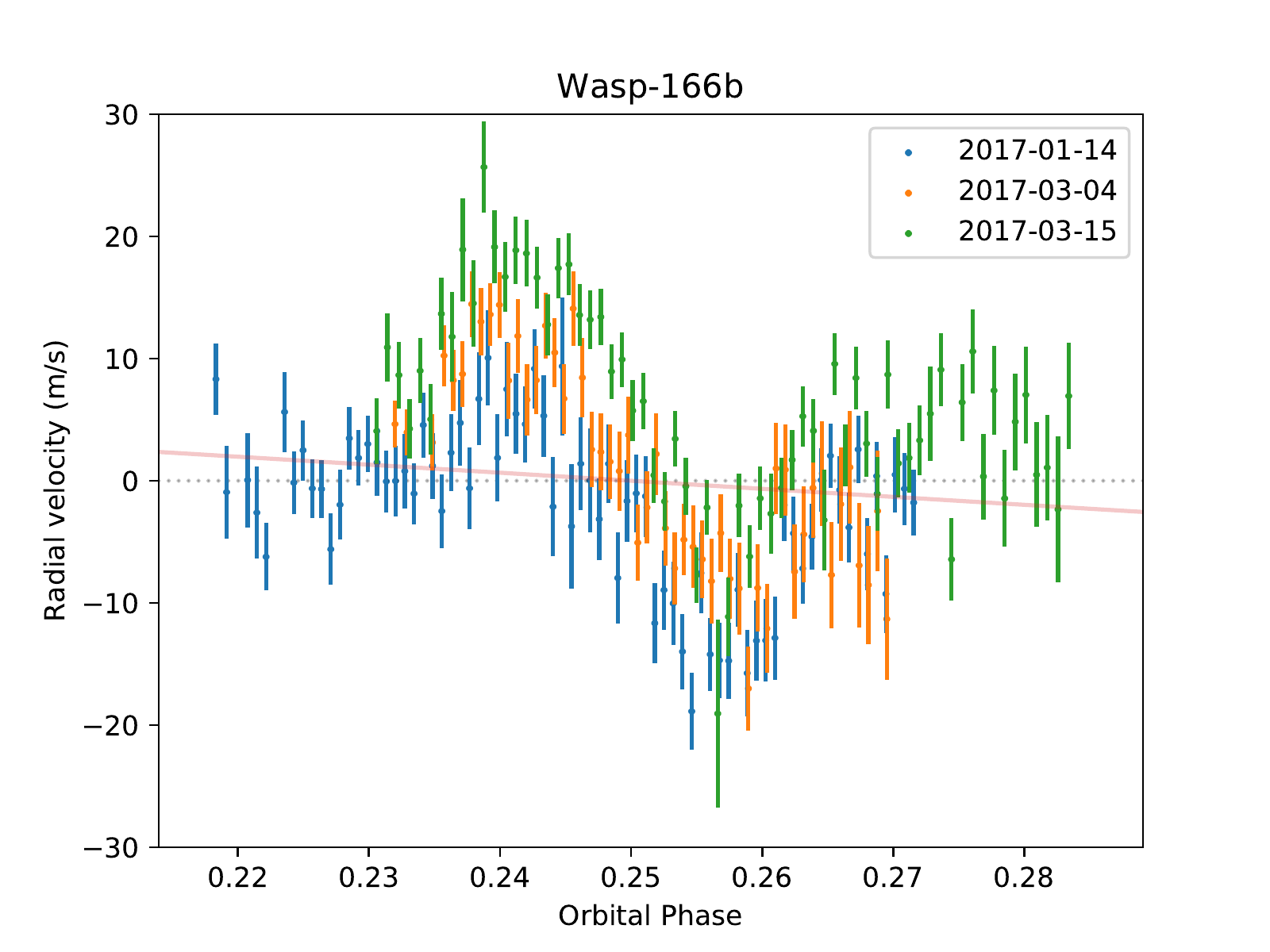}\\
     \includegraphics[width=0.45\hsize]{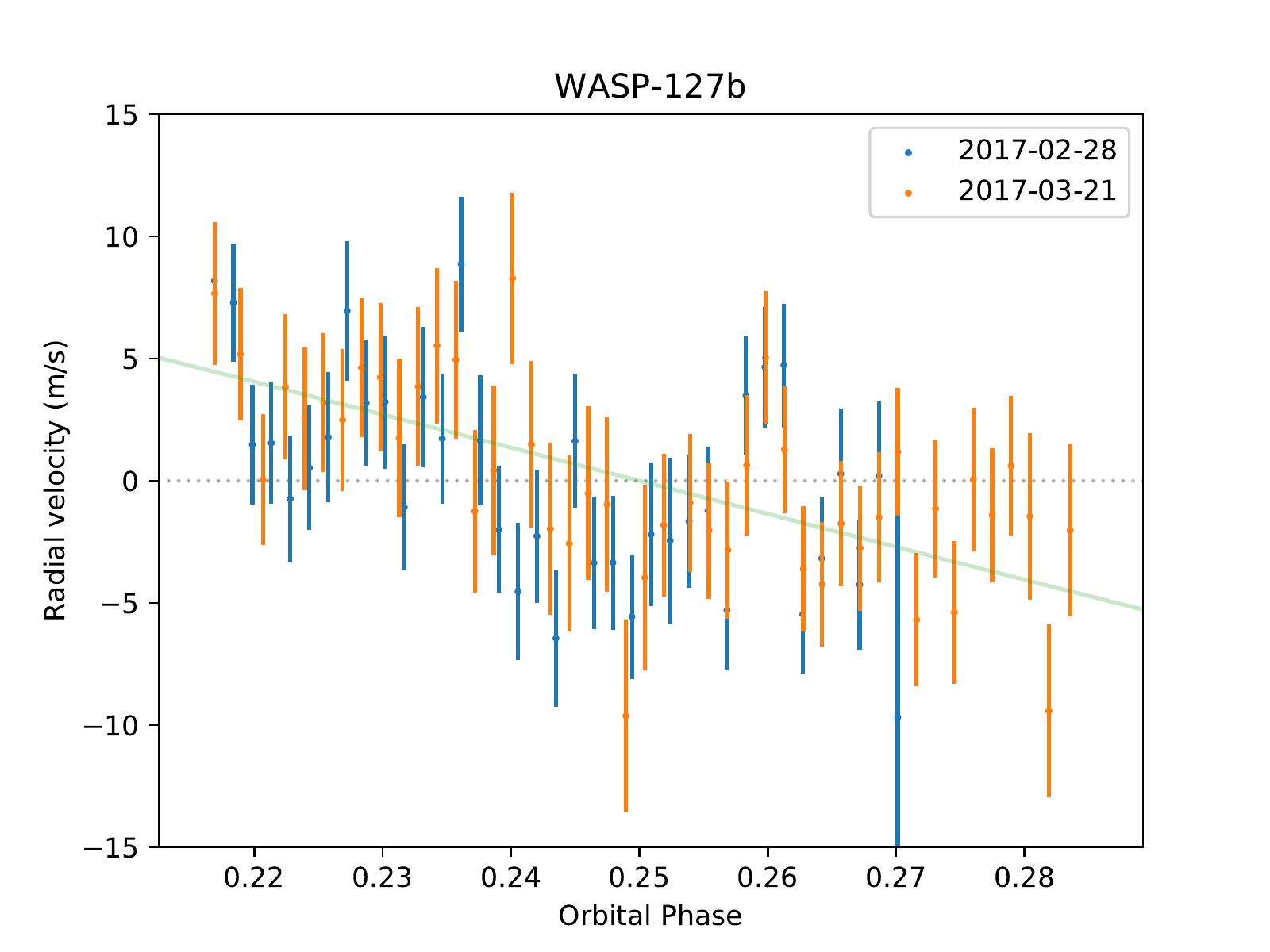}
    \includegraphics[width=0.45\hsize]{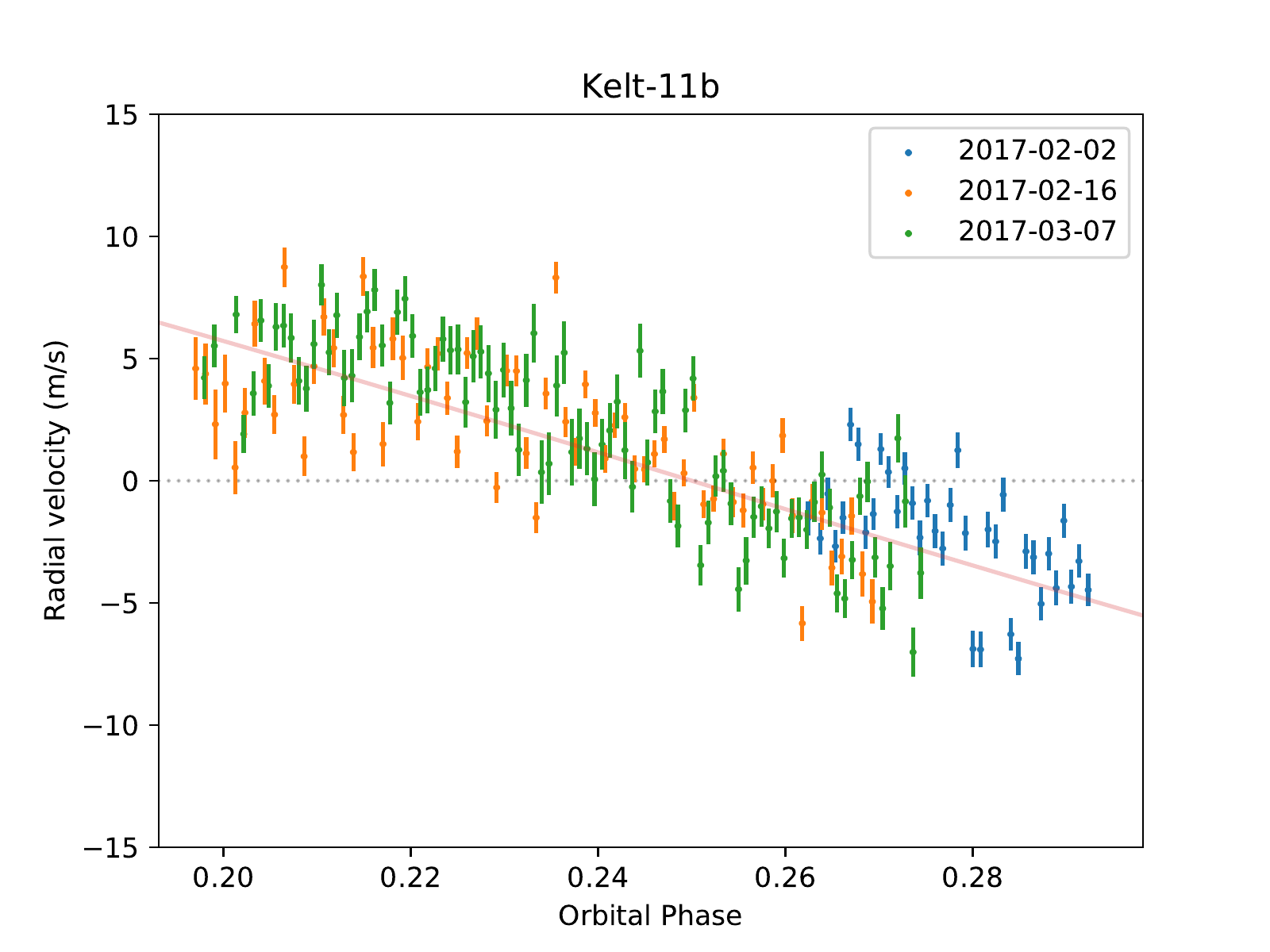}
     \caption{The radial velocities of our targets have been measured with HARPS: WASP-76 (top left), WASP-166 (top right), WASP-127 (bottom left) and KELT-11 (bottom right). There is some apparent offset between various epochs for the same target, which may be due to some instrumental shift or could be caused by the planet crossing star spots or faculae regions. In each case, the orbital motion is also indicated with the dashed line. Any deviation from this is due to the Rossiter--McLaughlin effect, which is clearly in all but one of our targets.}
         \label{fig:RM}
   \end{figure*}


  \begin{figure*}[htbp]
  \centering
   \includegraphics[width=0.45\hsize]{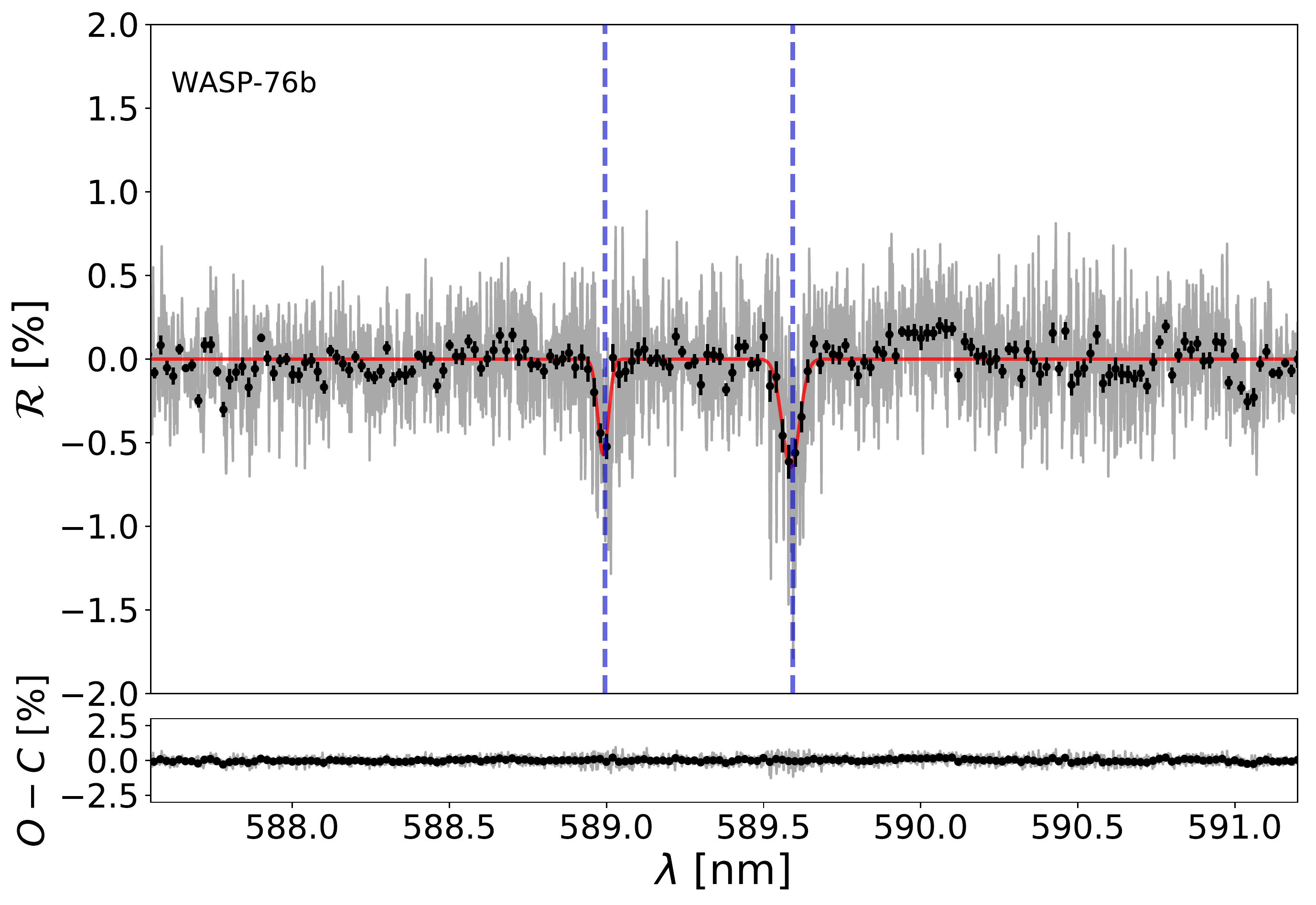}
    \includegraphics[width=0.45\hsize]{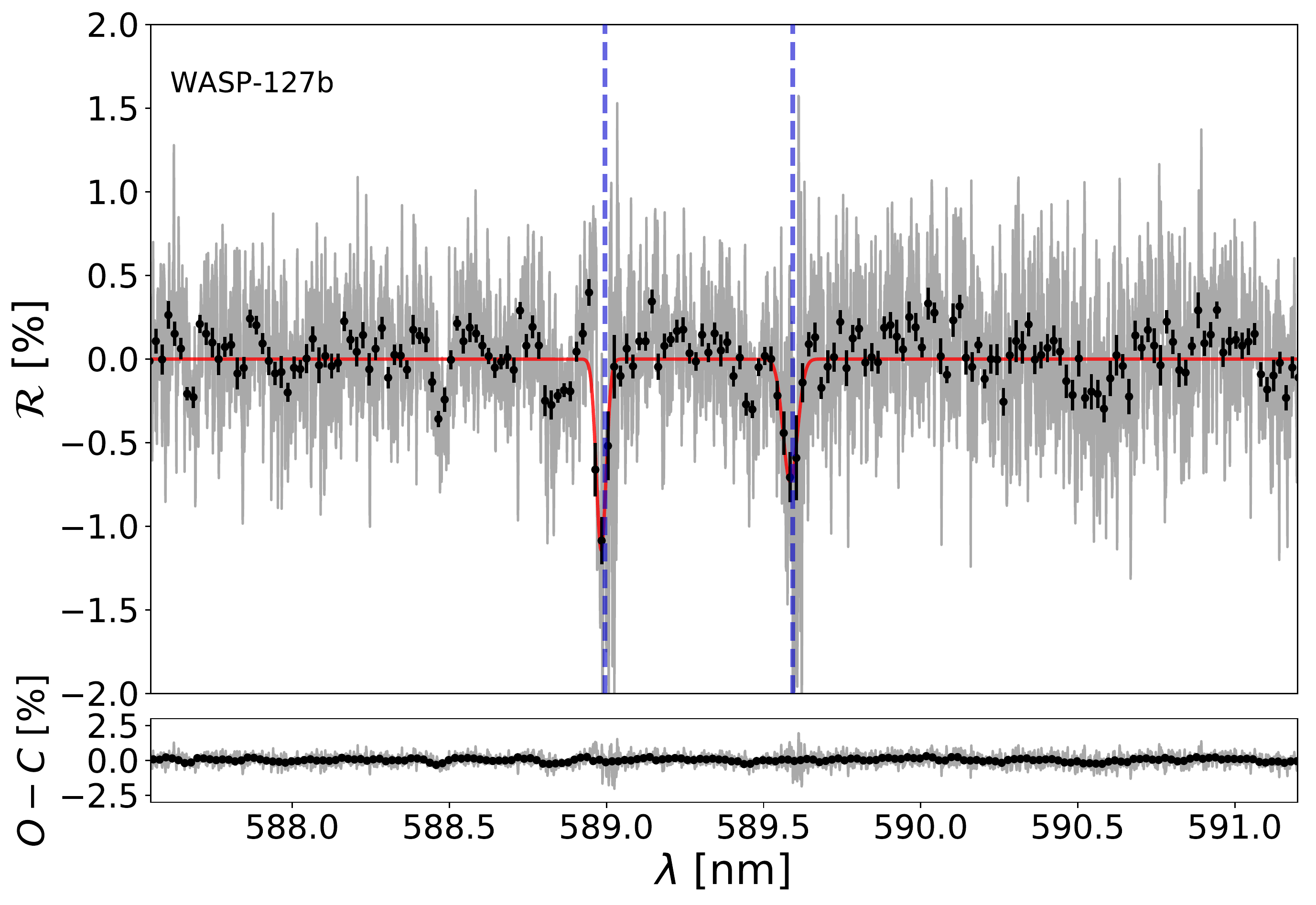}\\
    \includegraphics[width=0.45\hsize]{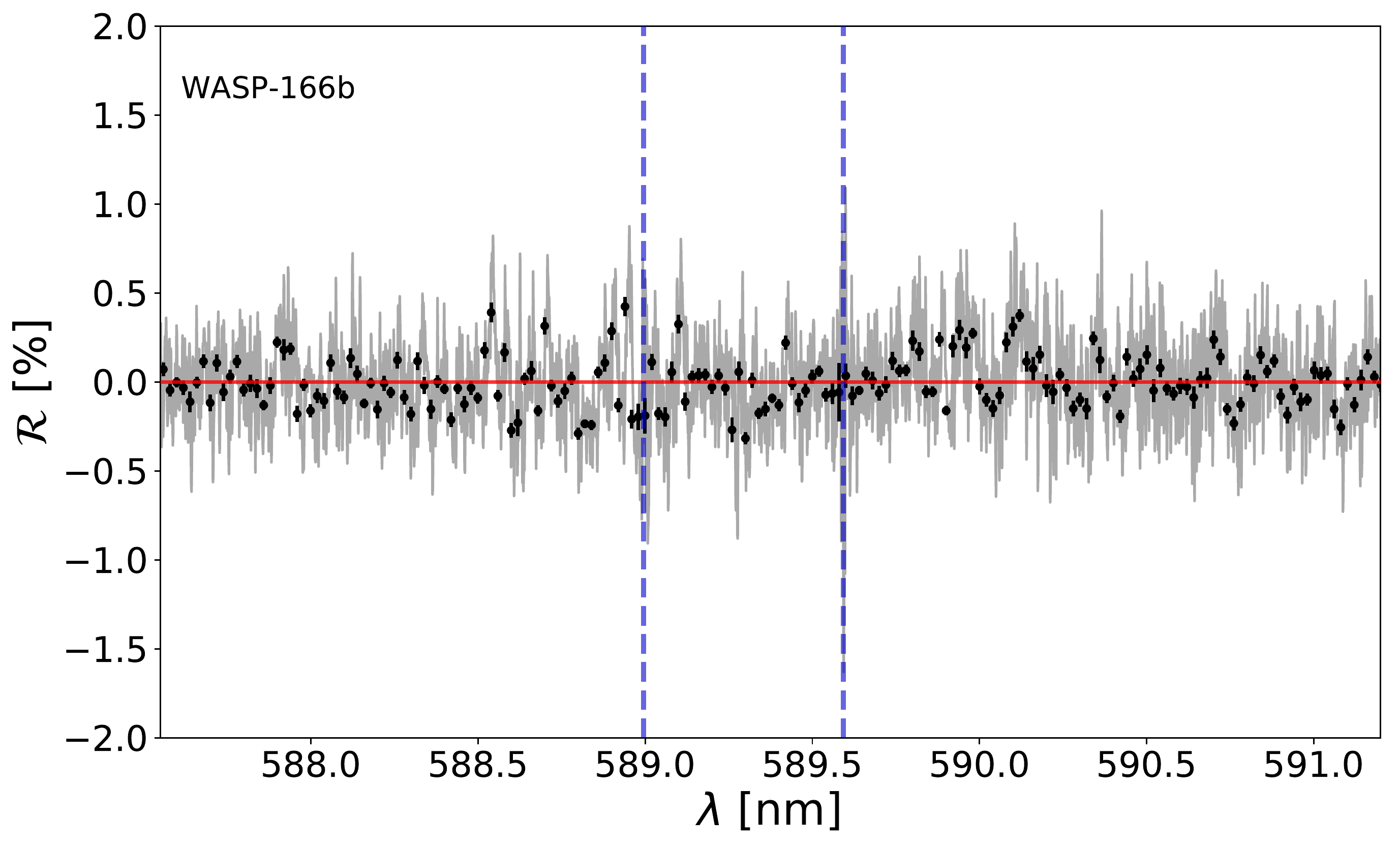}
    \includegraphics[width=0.45\hsize]{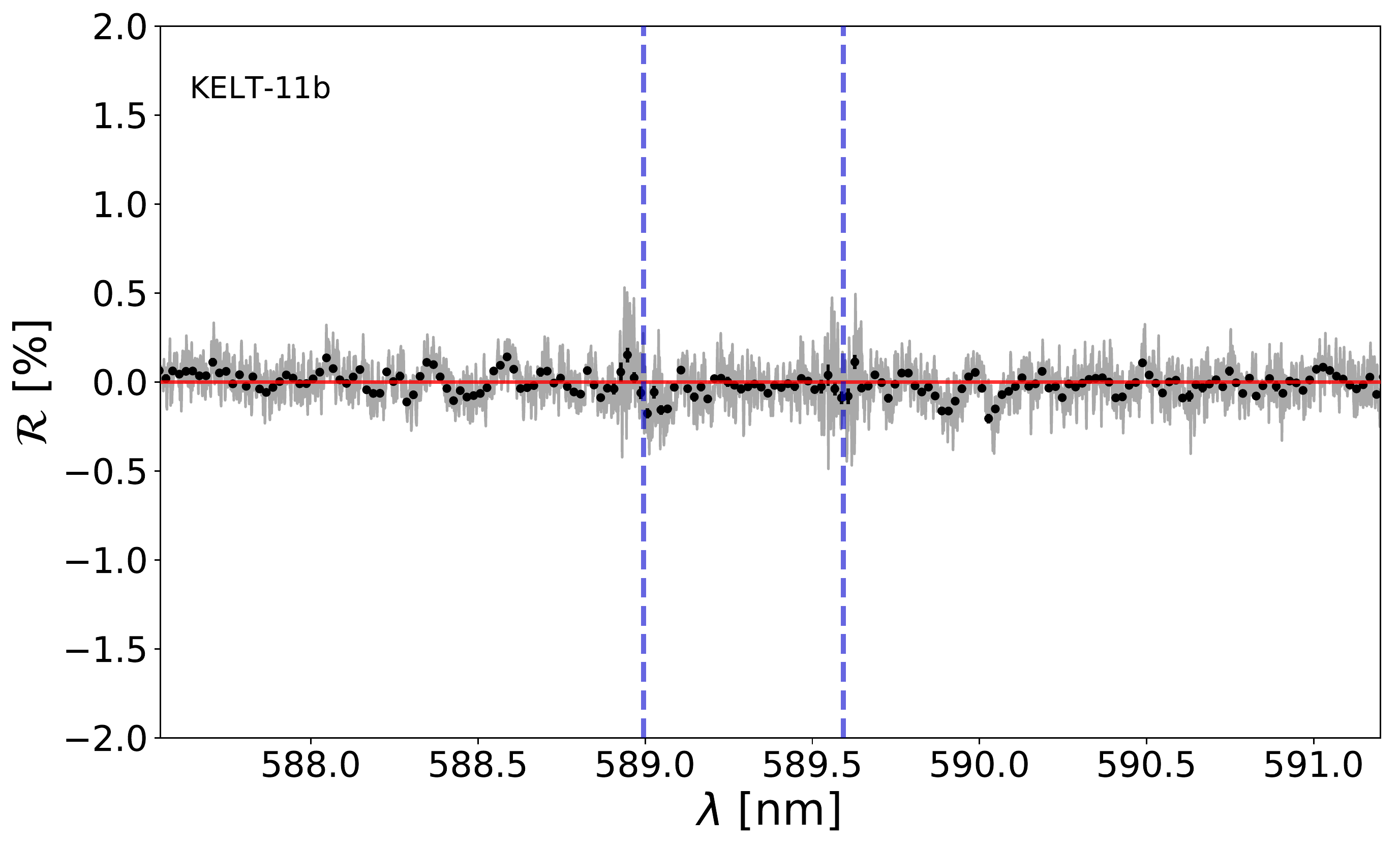}
     \caption{The spectral region around sodium doublet for our four targets, removing the stellar signature. For WASP-76b (top left) and WASP-127b (top right), sodium is clearly detected, while for WASP166-b (bottom left) and KELT-11b (bottom right), nothing is evident. The blue dashed lines indicate the position of the Na D2 and D1 lines.}
        \label{fig:na}
   \end{figure*}

\section{Results}\label{r}

Each data set was analyzed for Na D, H$_{\rm \alpha}$, H$_{\rm \beta}$ to see the absorption in the exo-atmosphere and for Ca I (612.2217 nm and 616.2173 nm) and Mg~I (518.3604 nm) to monitor the stellar activity. 
The results are described for each star separately.\\

\begin{table}
\caption{Parameters fits for the NaD lines in WASP-76b and WASP-127b.}             
\label{t:w76}      
\centering                          
\begin{tabular}{c c c}        
\hline\hline                 
Line & D1 & D2  \\    
\hline                        
\multicolumn{3}{c}{} \\
& \multicolumn{2}{c}{WASP-76b} \\
Depth ($\%$) & 0.648$\pm$0.068     & 0.57$\pm$0.08     \\      
FWHM (nm)    & 0.0650$\pm$0.0078 & 0.0400$\pm$0.0065 \\
\multicolumn{3}{c}{} \\
& \multicolumn{2}{c}{WASP-127b} \\
Depth ($\%$) & 0.735$\pm$0.088   & 1.144$\pm$0.270  \\      
FWHM (nm)    & 0.0537$\pm$0.0075 & 0.0367$\pm$0.0097 \\
 
\hline                                   
\end{tabular}
\end{table}

\noindent {\bf WASP-76b}:


The scale height of the atmosphere of WASP-76b is estimated to be about 1212 km, making it an excellent system for the analysis by transmission spectroscopy \citep{kabath}. This exoplanet was also suggested as one of the primary targets for JWST observations \citep{moliere}. Fig.\ref{fig:na} presents the final divided in/out transit spectra corrected for planetary radial velocity, displaying clearly visible planetary sodium (Na D lines) absorption. The RMS of the Na D region is 0.0027 and the RMS of the binned spectrum  by 20 is 0.0013. The sodium line depths are $0.65\,\%$ and $0.57\,\%$, respectively, therefore, corresponding to 9.5\,$\sigma$ and 7.1\,$\sigma$ significance. This is the first detection of sodium in the upper atmosphere of WASP-76b. Details of the detected peaks are summarized in Table \ref{t:w76}. 

Figure \ref{w76bact} shows the non-significant detections in Ca I and Mg I lines (RMS of the data points is 0.0007), showing that the stellar activity is low and any planetary atmosphere sodium (Na D lines) detection should be significant and not influenced by the star. Furthermore, the star WASP-76 is a slow rotator as reported by \cite{brown}. 

The D1 line is apparently stronger than the D2 line, which would imply different mixing ratios and compositions of the atmospheres of different planets because all other detections reported were with stronger D2 line. The explanation for the strength of the lines can be found in the literature for other Na lines in redder regions \citep{civis}, but such a measurement is missing in the optical. \cite{civis} present a laboratory spectrum of sodium lines demonstrating
that different atomic states are shaping the spectral lines and their strengths.
Furthermore, \cite{slanger} measured the D2/D1 ratio in the Earth's
atmosphere and found it to be varying between 1.2-1.8. They explained that the variability of the
ratio is originating from a competition between Oxygen reacting with
NaO($A^{3}\Sigma^{+}$), produced from the reaction of sodium with Oxygen. In other words, the line strength and the mixing ratio depend on the composition of  the surrounding
atmosphere and it can even vary over time.


In the course of the review process of this paper, an independent detection of the sodium in the atmosphere of WASP-76b planet by \citet{seidel} was reported for the same HARPS data set. Their results are in agreement with ours, confirming the methodology. We note, however, that our FWHM of the D2 line is about 30\,\% smaller compared to their values.\\

\noindent {\bf WASP-127b}:

WASP-127b is a heavily bloated super-Neptune. Its scale height is $2500\pm400$ km, making it another ideal target for transmission spectroscopy. Indeed, a cloudless sky was observed with many features such as sodium, potassium and lithium detected and hints of TiO absorption were reported with GTC instrumentation \citep{palle,palle2}. 

Fig. \ref{fig:na} presents the final divided in/out transit spectrum. The RMS of the data is 0.0042 and by binning by 20 points we reach a final RMS of 0.0019. We were able to confirm only sodium (Na D lines) with a high significance of 4.2-$\sigma$ (D2) and 8.33-$\sigma$ (D1), respectively. We did not detect lithium (670.8~nm), which is most likely due to aperture differences between the GTC (10\,m) and the 3.6\,m telescope that hosts HARPS, while as mentioned earlier, potassium (766.5 \& 769.9 nm) is out of range of HARPS. However, our confirmation of sodium is unique because this is one of the first proofs for observations using different instruments as well as two different methods. \\
The stellar activity of WASP-127 star can be monitored with the Ca I and Mg I lines. We were not able to see any activity in these lines, as shown in Fig.~\ref{w76bact}. \\


%
%

%




\noindent {\bf WASP-166b}:

 WASP-166b is a bloated super-Neptune with a scale height of 1103 km. Fig. \ref{fig:na} presents the final divided in/out transit spectrum of all regions of interest. The RMS of the data is 0.0025 and, after binning by 20 points, 0.0014. However, we are not able to detect any feature in the sodium (Na D) lines. Therefore, we can only put an upper limit of 0.14\,$\%$ on the detection.

Fig. \ref{w76bact} displays no signatures in the Mg I and Ca I lines. Therefore, we can assume that the star is not active. \\


%

\noindent {\bf KELT-11b}:


The scale height of KELT-11b is about 2500 km and since the planetary atmosphere is bloated, this is another excellent system for detection of exo-atmosphere \citep{kabath}.

Fig. \ref{fig:na} presents the final divided in/out transit spectrum of all regions of interest. The RMS of the data is 0.0012 (unbinned) and 0.0006 (binned by 20 points). The upper limit for a sodium (Na D lines) detection is therefore 0.06\,$\%$. This is a very stringent constraint on the presence of sodium in the atmosphere and possibly hints at the presence of clouds in the atmosphere of KELT-11b.

We checked the stellar activity on Ca I and Mg I lines. There was no significant detection of activity observed in our data (see Fig.\,\ref{w76bact}).




\subsection{Previous Na D detections}

We have investigated all the previous Na D line detections in exoplanets based on high-resolution spectroscopy. We have summarized them in Table \ref{t:otherdet}. As can be seen, WASP-76b is the first reported planet having potentially an absorption ratio of D2 to D1 lines less than one\footnote{One should note, however, that the deviation from unity is less than 1-$\sigma$, and we therefore do not claim much about this yet.}.\\
The unusual reported values make this planet an ideal target for future JWST follow-up to further clarify this discovery.

\begin{table}
\begin{threeparttable}
\caption{Selected high-resolution detections of sodium (Na D lines). The first column states the exoplanet with the source of the values. The second column shows the ratio of the sodium doublet absorption lines. The third column shows the ratio of the sodium doublet FWHMs. }           
\label{t:otherdet}

\centering                         
\begin{tabular}{c c c}       
\hline\hline                
Planet & $\rm{A_{D2}}/A_{D1}$ & $\rm{f_{D2}}/f_{D1}$  \\
\hline                    
HD\,189733b$^1$   &1.59 $\pm$\ 0.33 & 1 $\pm$\ 0.22\\
WASP-49b$^2$ & 1.09 $\pm$\ 0.47 & 1.91 $\pm$\ 0.83 \\   
MASCARA-2b$^3$ & 1.19 $\pm$\ 0.39& 0.79 $\pm$\ 0.31\\
WASP-17b$^4$  &1 $\pm$\ 0.66 & 1 $\pm$\ 0.57 \\
WASP-127b$^5$ & 1.56 $\pm$\ 0.41 &0.68$\pm$\ 0.20 \\
WASP-76b$^5$  & 0.88 $\pm$\ 0.15 & 0.62 $\pm$\ 0.12 \\
\hline                                   
\end{tabular}
\begin{tablenotes}
\item [1] \cite{wytt15} \item [2] \cite{wytt17} \item [3] \cite{casa18} \item [4] \cite{khala}\\ \item [5] This Paper
\end{tablenotes}
\end{threeparttable}
\end{table}
 
\section{Conclusions}\label{rmef}

We investigated a group of four inflated gas giant planets, among which two inflated hot-Jupiters and two super-Neptunes. All four targets are ideal targets for transmission spectroscopy, given their magnitudes and their large scale heights. We reported for the first time the detection of sodium in WASP-76b with a 7--9\,$\sigma$ significance.\\
We have been able to fully resolve the Na D lines. We report for the first time an absorption ratio of D2 to D1 lines less than one, although this is within 1-$\sigma$ of unity. If confirmed, this possibly hints at an unusual composition of the exo-atmosphere. \\ There was no hydrogen (H$\alpha$\ and\ H$\beta$\ lines) detected for WASP-76b. Another planet, WASP-127b was already characterized in previous studies and we could confirm the presence of sodium in the atmosphere. This is to our knowledge the first time such a detection is confirmed, and provides further credence to the transmission spectroscopy method given that two different instruments on two very different aperture telescopes (3.6-m vs. 10-m GTC)  provide similar results. For WASP-166b and KELT-11b, we were not able to detect any feature, despite having extremely precise data, thus hinting at cloudy skies or simply the absence of sodium and other elements. We note that non-detections are usually not reported, although they provide useful information as well. 

\acknowledgements
      The data presented here are available in the ESO archive under ESO Prog.IDs:\ 090.C-0540(F), 098.C-0304(A) and 0100.C-0750(A). 
PK and JZ would like to acknowledge the support from the GACR international grant 17-01752J. MS acknowledges financial support of Postdoc@MUNI project CZ$.02.2.69/0.0/0.0/16\_027/$ 0008360. This work benefited from support for mobility by the DAAD-18-08 programme, Project-ID: 57388159. We acknowledge the usage of IRAF and iSpec. We would like to
thank the anonymous referee for providing comments that helped to improve the manuscript.

\software{HARPS Data Reduction Software, iSpec (Blanco-Cuaresma et al. 2014), IRAF (Tody 1986)}

%
%

\end{document}